\documentclass[12 pt]{article}

\usepackage{amsmath,amstext,amsgen,amsbsy,amsopn,amsfonts,graphicx,amssymb,
rotating,overcite}

\begin{document}

\title{An Introduction to Quantum Chaos}

\author{Mason A. Porter \\ \\ Center for Applied Mathematics  \\ \\ Cornell 
University}

\date{July, 2001}

\maketitle

\begin{centering}
\section*{Abstract}
\end{centering}

\vspace{.1 in}

	Nonlinear dynamics (``chaos theory'') and quantum mechanics are two of
 the scientific triumphs of the 20th century.  The former lies at the heart of 
the modern interdisciplinary approach to science, whereas the latter has 
revolutionized physics.  Both chaos theory and quantum mechanics have achieved
 a fairly large level of glamour in the eyes of the general public.  The study
 of quantum chaos encompasses the application of dynamical systems theory in 
the quantum regime.  In the present article, we give a brief review of the 
origin and fundamentals of both quantum mechanics and nonlinear dynamics.  We 
recount the birth of dynamical systems theory and contrast chaotic motion with
 integrable motion.  We similarly recall the transition from classical to 
quantum mechanics and discuss the origin of the latter.  We then consider the 
interplay between nonlinear dynamics and quantum mechanics via a 
classification and explanation of the three types of quantum chaos.  We 
include several recent results in this discussion.

\vspace{.3 in}

\subsection*{MSC NOS 37K55, 35Q55}

\begin{centering}
\section{Introduction}
\end{centering}

	``In the beginning, there was Chaos.''  These words, though somewhat 
pretentious, serve as a microcosm for the excitement that has been engendered 
by nonlinear dynamics, which is often called \begin{itshape}chaos 
theory\end{itshape} among general audiences.  Nonlinear dynamics is at the 
heart of the modern interdisciplinary approach to science.  Many people, 
however, appreciate this subject only in a very limited sense.  Chaos theory 
was annointed a glamorous field of study when James Gleick's introduction to 
it appeared in print in 1987.\cite{gleick}  References to chaos are prevalent 
in popular culture.  It has been mentioned, for example, in 
\begin{itshape}Jurassic Park\end{itshape}, the movie 
\begin{itshape}Pi\end{itshape}, and an episode of 
\begin{itshape}The Simpsons\end{itshape}.  In such references, it is often 
grossly misapplied, demonstrating that although chaos is glamorized in popular
 culture, it is not really understood by the general public.  People outside 
the scientific community are far more aware of fractals and the pretty 
pictures that can be created with them than with the analysis behind them and 
the fact that such behavior serves as a good model for systems in several 
scientific fields, including physics, chemistry, biology, economics, and 
geology.\cite{strogatz}

	Quantum mechanics has achieved a similar lofty status in the public 
eye.  There have been references to it in countless movies, magazines, and 
television shows.  It has even encroached upon the English language.  Like 
dynamical systems theory, however, it is grossly misunderstood by public 
audiences.  The term \begin{itshape}quantum leap\end{itshape}, for example, 
refers to a \begin{itshape}very large\end{itshape} change, even though the 
quantum regime encompasses quantities that are so tiny that one cannot 
properly analyze the behavior they describe as part of a continuum.  Indeed, 
the quantal regime is one of small jumps rather than large ones.

	One of the goals of studying quantum chaos is to combine the paradigms
 encompassed by nonlinear dynamics and quantum mechanics into one coherent 
theory describing regimes in which both theories are relevant.  Unlike the two
 separate concepts, this notion is not well-developed.  Additionally, quantum 
chaos is virtually unknown to the general public, despite the fact that this 
subject seeks to reconcile two objects of the world's fascination.  In 
scientific circles, the notion of quantum chaos \begin{itshape}is\end{itshape}
 well-known, but its facets are not understood as well as its two underlying 
theories.  The purpose of the present paper is to remedy this situation by 
providing an introduction to quantum chaos as well as a brief survey of some 
of the prevalent ideas in this area of research.  Toward this end, we 
introduce some of the fundamental concepts of quantum mechanics and nonlinear 
dynamics before attempting to marry these two fields.  We contrast chaotic 
behavior with integrable motion and discuss how dynamical systems theory arose
 from the study of celestial mechanics.  Similarly, we contrast classical and 
quantum mechanics and then recount the origin of the latter.  Finally, we 
present an introduction to quantum chaos that includes a classification of its
 types, a survey of recent results, and an attempt to explain what quantum 
chaotic behavior actually represents.

\begin{centering}
\section{Order and Chaos}
\end{centering}

\vspace{-.1 in}

	One of the hallmarks of nonlinear dynamics is the concept of 
equilibria, which helps characterize a system's behavior--especially its 
long-term motion.  There are numerous types of equilibrium behavior that can 
occur in continuous dynamical systems, but such long-time behavior is 
restricted by the number of degrees-of-freedom (that is, by the 
dimensionality) of the system.  In other words, one ignores the transient 
behavior of a dynamical system and only considers the limiting behavior as $t 
\longrightarrow \pm \infty$.  In dissipative systems, one considers the basins
 of attraction and repulsion of a given dynamical system.  The extent of the 
possible complexity of a dynamical system's attracting and repelling sets is 
determined by the dimension of the system.\cite{strogatz,wiggins,gucken}  
Hamiltonian systems (which are conservative) do not possess global attractors 
or repellors, but their dynamics also becomes more complex as their 
dimensionality increases.

	A one-dimensional system may be described by a single (unforced) 
ordinary differential equation (ODE) of first order.  Its phase space is a 
line.  All solutions must either approach a steady state or blow up, because 
the topology of the phase line implies that all equilibrium points separate it
 into two distinct regions.  Moreover, any observed blow-up must be monotonic;
 there cannot be any spiraling or other complex behavior.  

\begin{figure}[htb] 
	\begin{centering}
		\leavevmode
		\includegraphics[width = 2.5 in, height = 3 in]{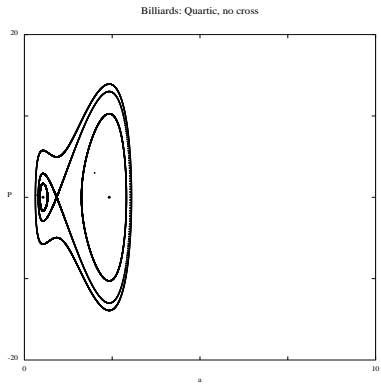}

\vspace{-.5 in}

		\caption{A separatrix that occurs in an integrable 
configuration of a vibrating quantum billiard in a double-well potential.  
Trajectories inside the separatrix behave qualitatively differently from those
 outside the separatrix.} \label{fig1}
	\end{centering}
\end{figure}

	Systems consisting of either an autonomous pair of first order ODEs or
 a single (unforced) second order ODE have two dimensions.  (A forcing term 
corresponds to a non-autonomity, which increases the dimensionality of the 
system when it is suspended into autonomous form.\cite{wiggins})  Such systems
 are aptly described by a phase plane.  Closed trajectories separate the plane
 into two parts, and if the qualitative behavior is different in those two 
regions of phase space, then the trajectories in question are known as 
\begin{itshape}separatrices\end{itshape} (see Figure \ref{fig1}).  Such 
separatrices are common in Hamiltonian systems, coming in the guise of 
\begin{itshape}homclinic\end{itshape} and 
\begin{itshape}heteroclinic\end{itshape} orbits.  Limiting behavior may 
include steady states, limit cycles, and blow up (which need not be 
monotonic).  Systems that are not Hamiltonian may thus exhibit various flavors
 of attractors and repellors.  

	The phase space of systems with $n \in [3,\infty)$ dimensions is 
embedded in $\mathbb{R}^n$.  In addition to the behavior that can show up in 
systems described in spaces of one or two dimensions, those with at least a 
third dimension may exhibit quasiperiodicity, chaotic (``strange'') attractors
 and repellers in addition to other manifestations of chaos such as 
ergodicity.  (Once again, attractors and repellors cannot occur in Hamiltonian
 systems, so one must distinguish chaotic behavior in those systems from that 
in dissipative and absorptive ones.  Such structures may prove to be relevant 
to the study of dissipative quantum chaos.\cite{haake,wall})  In many 
contexts, the concept of dimension is related to the number of 
\begin{itshape}degrees-of-freedom\end{itshape} (dof) of a system.  The number 
of \begin{itshape}dof\end{itshape} of a system is defined as the number of 
variables required to uniquely specify its orientation and position in 
physical space.\cite{kinetic}  This corresponds to the number of directions
 the system may move in configuration space.  For example, an unconstrained 
particle in open space may move in three different directions.  A (holonomic) 
system with $k$ degrees-of-freedom has a $2k$-dimensional phase space.  We 
will not treat non-holonomic systems\cite{goldstein,ms} in the present work.  
Such systems have velocity spaces of lower dimensionality than their 
configuration spaces so that a $k$ \begin{itshape}dof\end{itshape} 
nonholonomic system has a phase space of dimension $n < 2k$.  
\begin{itshape}Hamiltonian\end{itshape} systems are holonomic and 
conservative.  They may behave chaotically as well (as long as they possess at
 least two degrees-of-freedom), although their brand of chaos is somewhat 
different from that in other types of systems.

	Let us now compare periodic, quasiperiodic, and chaotic behavior.  To
 contrast the former two, consider the arcade game 
\begin{itshape}Asteroids\end{itshape}.  When the ship that the player controls
 flies off the screen on one side, it appears at the corresponding point on 
the other side, and the same is true of the top of the screen with respect to 
the bottom.  In other words, the game's playing field is a 2-torus, which is 
properly embedded in 3-dimensional space (which is why quasi-periodic behavior
 is possible).  More precisely, consider the following vector field on the 
torus:
\begin{gather}
	\dot{x} = 1, \notag \\   
	\dot{y}	= \omega.  \label{quasi}
\end{gather}
If $\omega$ is rational, the line in phase space eventually reaches its 
initial point so that it is periodic, whereas if $\omega$ is irrational, the 
line is quasiperiodic, approaching every point on the torus arbitrarily 
closely.  Another contrast between periodic and quasiperiodic motion occurs 
with planetary motion.  According to Kepler's First Law, each planet's orbit 
is an ellipse with the Sun at one of its foci.\cite{diacu}  A better 
model, however, is one in which planetary motion is described by precessional 
ellipses.  The latter motion is a quasiperiodic analog of the elliptical 
motion that describes the celestial body at any given instant.  The 
quasiperiodicity comes from the fact that the properties (eccentricity, angle 
of inclination, etc.) of the ellipse that describe the instantaneous motion 
change gradually over time.  Interaction with the other planets, in fact, 
leads to this evolution of the orbital parameters.  (Directly considering such
 time evolution is a way to incorporate perturbations due to the other planets
 in the solar system as a small perturbation of the two-body problem.)

	In addition to the theoretical distinction between periodicity and 
quasiperiodicity, there is an issue as to whether one can actually observe 
this difference.  (Equivalently, can nature tell if a number is rational or 
irrational in this context?)  Every irrational number can be approximated 
arbitrarily closely as a rational one,\cite{analysis} a fact that is very 
important for numerical simulations.  If one considers a single time series in
 Fourier space, one cannot in principle tell the difference between periodic 
and quasiperiodic motion, because computers approximate every irrational 
number as a rational one.  (Thus, the computer indicates that the result is 
periodic--though the period might be very long.)  However, one 
\begin{itshape}can\end{itshape} distinguish periodicity from quasiperiodicity 
based on the variation of a parameter if one computes multiple time-series 
plots.  For periodic motion, one observes that the ratio of given frequencies 
and higher harmonics remains constant, whereas this ratio varies across 
different time series in the quasiperiodic case.  On a computer, there is 
really no such thing as quasiperiodicity simply because every number is 
rational.  From a practical standpoint, however, one can tell the difference 
between perioic and quasiperiodic behavior as long as the reduced form of the 
rational number contains reasonably small integers (that is, as long as the 
period is not too long).  One must nevertheless be careful when doing 
dynamical computer simulations, as it may not always be easy to distinguish a 
quasiperiodic orbit and a periodic orbit with a very long period.

	We are now ready to contrast quasiperiodic and chaotic motion.  For 
motion to be chaotic, it must satisfy three properties: boundedness, infinite 
recurrence, and sensitive dependence on initial conditions.  The first 
property simply means that one can find a ball of sufficiently large radius 
that contains the chaotic attractor.  The second one implies that if one 
considers an arbitrarily small neighborhood about the initial point of a 
trajectory, it will return to the neighborhood infinitely many times.  The 
last property means that two trajectories that emanate arbitrarily closely 
diverge from each other at an exponential rate.  (That is, the trajectories 
are characterized by a positive Lyapunov exponent.)  Quasiperiodic motion 
satisfies the first two properties but does not satisfy the third.  Two nearby
 quasiperiodic trajectories remain ``close'' to each other in the sense that 
they only diverge linearly.

\begin{figure}[htb] 
	\begin{centering}
		\leavevmode
		\includegraphics[width = 2.5 in, height = 3 in]{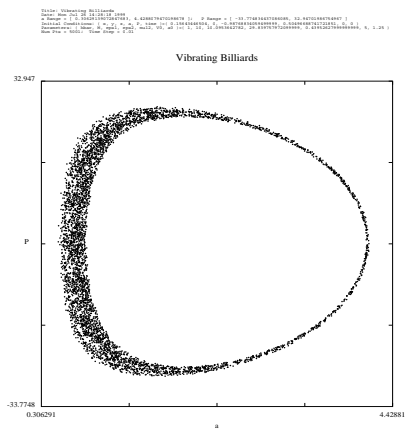}

\vspace{-.5 in}

		\caption{An example of hard Hamiltonian chaos.} \label{fig2}
	\end{centering}
\end{figure}

\begin{figure}[htb] 
	\begin{centering}
		\leavevmode
		\includegraphics[width = 2.5 in, height = 3 in]{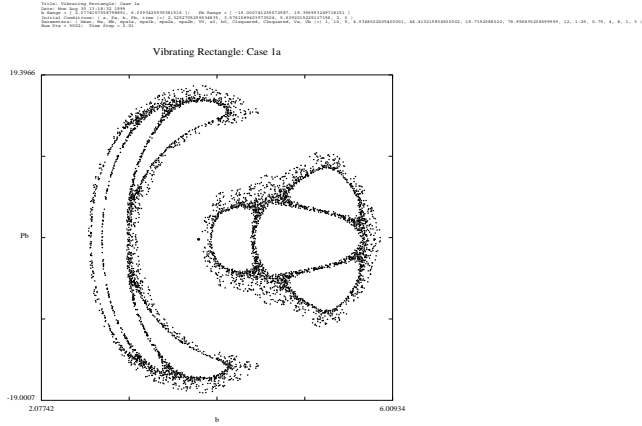}

\vspace{-.5 in}

		\caption{An example of soft Hamiltonian chaos.} \label{fig3}
	\end{centering}
\end{figure}

	One can also distinguish several different types of chaos.  There is 
Hamiltonian chaos as well as chaotic behavior in dissipative and absorptive 
systems.  Both classes of systems exhibit numerous flavors of chaos.  In the 
former, one can observe stochasticity (Figure \ref{fig2}), in which a 
Poincar\'e map displays a bounded set within which there is complete disorder.
  One can also observe so-called ``soft chaos'' or ``local chaos'' (Figure 
\ref{fig3}), in which there is some fuzziness near otherwise well-defined 
curves.  If the curves are completely well-defined in a Poincar\'e map--that 
is, they lack fuzziness--they describe quasiperiodic behavior, and if one has 
a collection of dots rather than a curve, then the depicted motion is 
periodic.  One can also observe chaotic behavior that falls between these two 
cases.  This quasiperiodic route to chaos is described by KAM 
theory.\cite{gucken,wiggins}

	Finally, one can extend dynamical systems theory to partial 
differential equations (PDEs).  A PDE, which may be treated as an infinite set
 of coupled ODEs, has infinitely many degrees-of-freedom.\cite{infinite}  Such
 systems may exhibit chaotic behavior in both spatial and temporal variables. 
 (The chaotic behavior discussed above is temporal chaos.)  Possible 
manifestations of such spatio-temporal chaos include a spiral wave route to 
chaos anologous to the period doubling route to temporal chaos described by 
the Feigenbaum sequence.\cite{devaney}  Research concerning such 
spatio-temporal complexity is quite active.

\vspace{-.1 in}

\begin{centering}
\section{The Origin of Dynamical Systems Theory}
\end{centering}

	The historical evolution of dynamical systems began with the advent of
 celestial mechanics.\cite{diacu,celest}  The orbits of the planets were 
described as the ``Music of the Spheres,'' and the solar system was treated as
 something both magical and mystical.  Considered strong evidence of Divine 
creation, many scientists set out to explain this extraordinary natural 
symmetry.  Nicholas Copernicus (1574--1642), an early celestial mechanician, 
waited until he was on his deathbed to publish his heliocentric theory because
 he knew that it would be considered blasphemy.  Other pioneers in celestial 
mechanics, such as Galileo Galilei (1564--1642) and Johannes Kepler 
(1571--1630), also had to be careful with that they published on the subject. 
 (Kepler studied celestial bodies in part to explain their divine symmetries, 
whereas Galileo's work was treated quite harshly by the Church.)  Sir Isaac 
Newton (1642--1727) used Kepler's three laws of planetary motion and three laws
 of his own to derive the inverse square law of gravitational attraction.  
Essentially, Newton solved the (unperturbed) 2-body problem.

	A natural progression of these results was the attempt to derive a 
solution to the $n$-body problem.  According to the intellectual climate prior
 to the 20th century, the universe was a giant orrery that could be completely
 solved.  Nobody had yet succeeded in solving the $n$-body problem, but surely
 somebody would if given the proper motivation.  Scientists such as Pierre 
Simon de Laplace (1749--1827), Joseph Louis Lagrange (1736--1813), Sim\'eon 
Denis Poisson (1781--1840), and Spiru Haretu (1851--1912) attempted to analyze
 the stability of the solar system by examining the $n$-body problem, but 
their results were inconclusive.  Moreover, Haretu showed that the methods 
they were using were doomed to indeterminacy.  The problem was of such a 
magnitude that the development of new methods was essential to its resolution.

	In Volume 7 (1885--86) of 
\begin{itshape}Acta Mathematica\end{itshape} was an announcement that King 
Oscar II of Sweden and Norway would award a substantial prize and medal to the
 first person to obtain a global general solution to the $n$-body problem. 
Jules Henri Poincar\'e (1854--1912) had developed new techniques for studying 
differential equations, and he felt that these would provide a good intuitive 
basis for his attempt at this solution.  After more than two years of study, 
the nature of the situation began to take shape.  One of the problem's secrets
 was revealed by the special case $n = 3$.  Poincar\'e proved that there did 
not exist uniform first integrals other than the one that had already been 
found.  This showed that the 3-body problem could not be solved quantitatively
 by Hamiltonian dynamics (by using first integrals to reduce the problem to a 
solvable one of lower dimension), as even the restricted three body problem
needed two degrees-of-freedom to describe it fully.  The $n$-body problem was 
thus considerably more difficult than anyone had realized.  Mathematicians 
would have to change the way they treated systems of this sort.  They could no
 longer rely on quantitative methods to study the universe, a fact that 
countered the prevailing philosophy.  The presence of chaotic behavior showed 
that determinism did not imply accurate prediction, because even a small 
perturbation of the initial data of a problem could cause an arbitrarily large
 alteration in the behavior of its solution.

	From Poincar\'e's discovery arose dynamical systems theory.  In 
addition to showing that the $n$-body problem could not be solved 
analytically, Poincar\'e discovered the first manifestation of chaotic 
behavior in the form of homoclinic tangles.  Poincar\'e's work served as the 
foundation for that of several mathematicians and scientists--including such 
luminaries as George Birkhoff, Stephen Smale, Andrei Kolmogorov, Vladimir 
Arnold, and J\"urgen Moser--and the theory and methods he originated now hold 
prominent places in mathematics, science, and popular culture.

\begin{centering}
\section{Classical and Quantum Physics}
\end{centering}

\vspace{-.1 in}

	When introducing quantum mechanics to his students, Richard Feynman 
called it ``the description of the behavior of matter in all its details
 and, in particular, of the happenings on an atomic scale.''\cite{feynman1}  
Objects on this scale behave like neither particles nor waves.  The quantum 
behavior of all atomic objects is the same; there are respects in which they 
behave like particles and other respects in which they behave like waves.  The 
reconciliation of this particle-wave duality of matter is at the heart of the 
transition from classical to quantum mechanics.

	In quantum mechanics, quantities such as position, momentum, and 
energy play roles as operators as well as variables (depending on the 
representation in use).  In classical mechanics, only the latter role is 
played.  Concomitant with this additional interpretation is the issue of 
compatibility of observable quantities.  Incompatible observables do not 
commute, and so there is an `uncertainty relation' between them.\cite{sakurai}
  In such relations, the more precisely one knows one quantity, the less 
precisely one can determine the other.  The canonical Heisenberg Uncertainty 
Principle expresses this phenomenon between position and momentum, a pair of 
complementary quantities.  From a mathematical point of view, this example of 
quantum-mechanical uncertainty is a consequence of the Fourier Integral 
Theorem.\cite{butkov}  (One must be careful in more general situations, as 
this result depends on the fact that position and momentum are Fourier 
transforms of each other.)

	Another contrast between classical and quantum physics is that the 
former has a continuous energy spectrum, whereas that of the latter 
is discrete.  This is perhaps best illustrated by comparing a classical 
oscillator with a quantum-mechanical one.  In both cases, one can find `normal
 modes' (`eigenfunctions', `wavefunctions') $\psi_n(r,t)$ and their associated
 eigenvalues (`eigenenergies').  One then considers an arbitrary superposition
 of these normal modes:
\begin{equation}
	\psi(r,t) = \sum_{n=0}^\infty c_n \psi_n(r,t). \label{mode}
\end{equation}
Equation (\ref{mode}) is traditionally interpreted differently in classical 
and quantum mechanics.  In the former, $\psi(r,t)$ is a linear combination of 
all the normal modes of an oscillator.  One can observe this superposition, 
for example, when conducting experiments with a string or a Slinky.  One such 
experiment is to demonstrate that `travelling waves' are a solution of the 
canonical wave equation by showing that they are one example of a 
superposition of normal modes.  In quantum mechanics, one can only observe a 
single wavefunction at one time.  Instead of being an expression for the 
degree of expression of a given normal mode, the coefficients $c_n$ in the 
eigenfunction expansion for $\psi(r,t)$ are instead interpreted as a measure 
of the likelihood that the $n$th wavefunction $\psi_n(r,t)$ manifests in a 
given experiment.  The superposition $\psi(r,t)$, then, represents an 
expectation of possible wavefunctionns rather than a linear combinations of 
observed ones as in classical mechanics.  

	In the above discussion, we were careful to indicate that we were 
referring to the discreteness of energy rather than of the eigenvalues some 
other quantum-mechanical operator, for which one may yet have a continuum.  
The momentum operator, for example, may admit such a continuum.  Additionally,
 the distinction in the preceeding paragraph is in many ways cosmetic, as 
classical Sturm-Liouville operators have discrete energy 
spectra.\cite{simmons}  Nevertheless, this distinction is a convenient one to 
use from a historical and expository perspective--it is extremely useful for 
elementary discussions of quantum mechanics.  In truth, however, the situation 
is subtler and more complicated, as there are important dynamical differences 
depending on whether a quantum system's spectrum is discrete, continuous, or 
contains regions with each property.  In fact, such differences lie at the 
heart of the search for genuine quantum chaos, because any candidate system 
that might exhibit such behavior must be (spatially) bounded and unforced with
 a spectrum that is \begin{itshape}not\end{itshape} discrete.  It must also be
 fully quantum and describe a finite number of particles.\cite{ilg}

	 The probabilistic interpretation of quantum mechanics highlights an 
important difference between classical and quantum physics.  In 
quantum-mechanical experiments, it is impossible to predict exactly what can 
happen in a given circumstance in the same sense as in classical experiments, 
as one can only observe a single mode at a time.  Instead, one predicts the 
probabilities of different events, which can then be measured by repeated 
experimentation.  However, this distinction between classical and quantum 
mechanics is, in some sense, cosmetic.  One can, for example, compute 
probabilities in classical mechanics---the key difference is that all 
observable quantities may be measured simultaneously so that the expressions 
that correspond to the off-diagonal elements in a quantum-mechanical matrix 
calculation must necessarily vanish.  This, moreover, is intimately related to
 what is perhaps the fundamental difference between the classical and quantum 
theories.  In classical physics, particles can be labeled by their position 
and velocity at a given time and their trajectories are thereby distinguished.
  In the quantum regime, however, particles do not have definite trajectories,
 so this distinction cannot be made.  Indeed, in any experiment, one can 
switch the labeling of two identicle particles without altering its outcome.

	Like dynamical systems theory, quantum mechanics has important 
philosophical implications.\cite{feynman1}  The Heisenberg Uncertainty 
Principle implies that making (highly non-perturbative) observations of a 
phenomenon affects the phenomenon itself.  This effect, moreover, cannot be 
minimized arbitrarily by altering experiments.  There is a minimum disturbance
 that one simply cannot avoid.  In classical physics, an observer is important
 only in a passive sense, whereas in quantum mechanics, the effects of an 
observation can be highly nontrivial.

\vspace{-.1 in}

\begin{centering}
\section{The Origin of Quantum Mechanics}
\end{centering}

	Now that we have highlighted several distinctions between quantum and 
classical mechanics, let us review its history.  The physics community was in 
turmoil as the 19th century faded into oblivion and the 20th century 
began.\cite{liboff}  Amidst this maelstrom lay the origin of quantum 
mechanics.  There were many experimental observations that were inexplicable 
according to the firmly grounded classical theory.  As with nonlinear 
dynamics, the theoretical answer to these questions required a new way of 
thinking.  At its very core lay subjective probability rather than objective 
determinism.  

	One of the aforementioned experimental observations was that light 
exhibits interference fringes and is therefore a wave phenomenon.  However, 
erroneous results are obtained if one attempts to explain the photoelectric 
effect using the postulate that light is a wave.  One finds that the energy of
 an emitted electron depends only on the frequency of the incident radiation 
rather than on its intensity as one might expect from the classical optics.  

	In 1901, Max Planck (1858--1947) observed blackbody radiation, showed 
that energy due to radiation could only exist in the form of discrete packets,
 and introduced his constant $\hbar$.  In formulating his explanation of 
radiation packets, Planck had to abandon the notion that the second law of 
thermodynamics was an absolute law of nature.  It was instead a statistical 
law.  In 1905, Albert Einstein (1879--1955) discovered the photoelectric 
effect.  Six years later, Ernest Rutherford (1871--1937) found that an atom 
has a positive central core surrounded by satellite electrons.  Such 
circulating (and hence accelerating) particles radiate energy and so---based 
on classical theory---one would expect the electron to collapse into the 
nucleus.  Why, then, does one not observe a burst of ultraviolet radiation 
emitted as an electron spirals into the nucleus?  Why, moreover, is the 
frequency spectrum of light emitted from an atom discrete rather than 
continuous?  Niels Bohr's (1885--1962) response, published in 1913, was his 
quantum theory of spectra.  

	In 1922, Arthur Compton (1892--1962) discovered that photons scattered
 off electrons.  Two years later, Wolfgang Pauli (1900--1958) published his 
famous Exclusion Principle, which states that there are no fermion states in 
which two or more particles share the same quantum numbers.  In 1925, Louis de
 Broglie (1892--1987) proposed that the wave-particle duality is a universal 
characteristic of particles in nature rather than just a phenomenon that is 
observed in light.\cite{merz}  He claimed that the wave nature of matter would
 become evident when the magnitude of Planck's constant $\hbar$ could not be 
ignored.  As a consequence, one could observe diffraction patterns from beams 
of particles other than photons.  

	Because matter exhibits aspects of both particles and waves, one must 
modify the tenets of classical physics.  Bohr stressed the need for 
reconciling this wave-particle duality, introducing his concept of 
``complementarity'' in 1927.  In order to accomplish this, quantum theory must
 account for the discreteness of certain physical properties that entered the 
realm of physics before Bohr's atomic model.  Normal modes, for example, are a
 quantized phenomenon from classical mechanics.  Classically, one considers a 
superposition of such modes to describe the motion of an oscillator, whereas 
in quantum mechanics one considers an expectation of `normal modes' (that is, 
wavefunctions) to describe the system of interest.  

	The previous year, Erwin Schr\"odinger (1887--1961) wrote an 
equation describing the wavefunction of a particle based on the laws of 
quantum mechanics he had discovered.\cite{feynman3}  This partial differential
 equation,
\begin{equation}
	i \hbar \frac{\partial \psi(r,t)}{\partial t} = H \psi(r,t),
\end{equation}
was similar to the classical equations that describe other types of waves such
 as those describing sound, light, and (especially) heat.  When quantum 
mechanics was first postulated, most scientsists studying it spent the 
majority of their effort attempting to solve the Schr\"odinger equation.  
Other physicists, including Max Born (1882--1970) and Paul Dirac (1902--1984),
 extended quantum mechanics further by incorporating phenomena such as spin.  

	In 1927, Werner Heisenberg (1901--1976) discovered his Uncertainty 
Principle, which implies that corresponding to a smaller error in the 
measurement of a particle's momentum must be a larger one in a simultaneous 
measurement of the particle's position (and vice versa).  For example, if one 
performs an identical experiment many times in which the position of an 
electron is measured (with a given momentum), then measurement of the position
 does not give an identical result for each experiment.  A consequence of 
Heisenberg's principle is that there are compatible variables that can be 
simultaneously measured and incompatible ones that cannot be.  The same year, 
Clinton Davisson (1881--1958) and Lester Germer conducted experiments on the 
wave properties of electrons and thereby demonstrated electron diffraction.  
Also in 1927, Max Born introduced a probabilistic interpretation of 
wavefunctions $\psi$.  He claimed that the quantity $\|\psi\|^2$ was properly 
treated as a probability density rather than just as the intensity of a wave. 
 Born's postulate is consistent with the results pertaining to the 
interference of electrons and photons.  The presence of particles leads to 
interference fringes, which is generated by the wavefunction.  At positions 
for which $\|\psi\|^2$ is large, the probability that the particle is found 
there is concomitantly large.  In the presence of many particles, their 
distribution is described by the probability density function $\|\psi\|^2$.

	Additional discoveries have molded quantum mechanics into its modern 
form.  In 1928, Dirac discovered a relativistic wave equation and predicted 
the existence of the positron.  As quantum mechanics continued to develop, 
physicists realized that there were many phenomena not directly encompassed in
 the Schr\"odinger equation, including electron spin and relativistic 
effects.\cite{feynman3}  The filling of these gaps led to relativistic quantum
 mechanics and quantum field theories.

\begin{centering}
\section{Making Sense Out of Quantum Chaos}
\end{centering}

\vspace{-.1 in}

	When given a mystery to solve, one seeks to find the simplest model 
possible that properly explains the unknown phenomenon.  In science as in 
life, it is beneficial to follow the dictums of Occam's Razor.  There will 
always be debates among scientists as to whether models contain extraneous 
elements or fail to offer satisfactory explanations, but the basis of 
simplicity in scientific modelling is one that is almost universally followed.
  A model must not only explain the desired phenomena, but it must be 
accessible to as many people as possible subject to that constraint.  The 
success of a model depends not only on its relation to reality but also on its
 practicality.  One can, in principle, incorporate every intricate detail when
 modelling phenomena, but this serves little purpose if it obscures what is 
important.  Additionally, one also strives to incorporate as many phenomena as
 possible in such a way as to provide a useful abstraction of the component of
 the universe in which one is interested.  This dichotomy, in fact, is 
practically an uncertainty principle in and of itself.  As one complexifies a 
model, it (in theory) describes reality more accurately, but it is also 
simultaneously more difficult to understand and analyze.  

	Dynamicists have yet to establish a consensus as to what types of 
behavior constitute quantum chaos.  One of the primary goals of the present 
paper is to offer a classification of the types of quantum chaos as well as an
 explanation of the behavior and analysis that characterizes each type.  In 
the present paper, we attempt to extend to the chaotic regime the 
well-established differences between classical and quantum mechanics.  
Mathematical objects such as wavefunction superpositions are interpreted 
differently in the two subjects even when the behavior is integrable, so we 
generalize such distinctions to the case in which a system behaves 
chaotically.  The differences between classical and quantum mechanics in the 
chaotic regime are a natural extension of those in the integrable regime, but 
the behavioral consequences of these differences can often be quite profound.

	Dynamical systems theory brought about a revolution in deterministic 
thinking in science just as quantum mechanics fomented a probabilistic 
revolution.  The study of quantum chaos is an attempt to marry these two 
abstract ways of thinking into a coherent whole in order to describe systems 
for which both quantum mechanics and nonlinear dynamics are relevant.  More 
specifically, the concept of quantum chaos is an attempt to extend the notions
 of classical Hamiltonian chaos to the quantum regime.\cite{gutz}  In the 
present paper, we review a classification scheme for quantum chaos and offer 
an exposition of the types of behavior described by the term `quantum 
chaos.'\cite{atomic}  There are three types of quantum chaotic behavior:  
``quantized chaos,'' ``semiquantum chaos,'' and genuine ``quantum chaos.''  
Our discussion of quantized chaos is influenced by those in 
\begin{itshape}Chaos in Atomic Physics\end{itshape}\cite{atomic} and 
\begin{itshape}Chaos in Classical and Quantum 
Mechanics\end{itshape}\cite{gutz}, and we mostly follow the work of Porter and
 Liboff\cite{sazim,nec,rect,bif,ksu} in our discussion of semiquantum chaos.  
We draw our discussion of true quantum chaos from that in \begin{itshape}Chaos
 in Classical in Atomic Physics.\end{itshape}

\vspace{-.1 in}

\begin{centering}
\subsection{Type I: Quantized Chaos}
\end{centering}

	Quantized chaos, also known as ``quantum chaology,'' is the most 
frequently studied form of quantum chaos.  This subject is primarily concerned
 with bounded autonomous systems with discrete spectra.  Quantized chaos 
concerns the quantization of classically chaotic systems, usually in the 
semi-classical ($\hbar \longrightarrow 0$) or high quantum-number regimes.  
One looks for signatures of classically chaotic systems on the quantum level. 
 In quantizing a chaotic system, one obtains a configuration that though not 
chaotic in a rigorous sense nevertheless behaves in a intrinsically different 
manner than an integrable system that has been similarly quantized.  
Nevertheless, the quantum dynamics of such systems are still affected in a 
fundamental manner by the fact that their classical counterparts are chaotic. 
 It has been shown, for example, that all atoms and molecules except the 
hydrogen atom (and related two-body atomic systems) exhibit chaos when treated 
classically.\cite{gutz,atomic}  The quantum dynamics of these systems are not 
rigorously chaotic, yet their associated waves and energies are strongly 
influenced by the underlying classical chaos.  The reason a quantum-mechanical
 system so obtained is not rigorously chaotic follows from the discrete nature
 of its energy spectrum.  In classical dynamics, chaotic behavior satisfied 
boundedness, infinite recurrence, and exponential sensitivity.  The properties
 of boundedness and infinite recurrence can be applied to the quantum regime 
in a well-defined manner.  One might worry that the discrete energy spectrum 
alters the topology so that one has to be careful about what it means for a 
``trajectory'' (a term we use loosely in the quantum regime) to recur in an 
infinitesimal neighborhood, but this is not actually a problem, because one 
can consider things in a different space, such as by taking Fourier transforms
 between the position and momentum bases.  The concept of exponential 
sensitity, however, cannot be used directly in the quantum regime, because of 
the so-called ``break time'' phenomenon.  In classical dynamics, exponential 
sensitivity refers to the exponentially fast separation of trajectories that 
began infinitesimally close to each other.  If one considers Lyapunov 
exponents in configuration space, one obtains a time series that increases 
linearly at first but then tapers off to be roughly constant as ``saturation''
 occurs.\cite{strogatz}  In a bounded system, two trajectories can only be 
separated by a finite distance.  In tangent spaces, however, this restriction 
is no longer present.  (The \begin{itshape}tangent space\end{itshape} at a 
point in a manifold is the set of all tangent vectors at that 
point.\cite{mta,ms})  Time-series plots of Lyapunov exponents in such spaces 
thus display a linear increase for all time.  In the quantum regime, however, 
one obtains saturation both in configuration space and in the tangent 
space.\cite{break}  The saturation in the tangent space referred to as the 
quantum \begin{itshape}break-time\end{itshape} phenomenon.  It is the reason 
that the concepts of Lyapunov exponents and exponential sensitivity break down
 in quantum physics.  Given that the systems under consideration satisfy 
boundedness and infinite recurrence rigorously and that there is still 
exponential sensitivity in some sense, it is not unreasonable to call these 
systems chaotic even though they are not rigorously so.  Nevertheless, we 
elect not do so, as it is insightful to distinguish this type of behavior from
 that which is rigorously chaotic.  The search for so-called genuine quantum 
chaos is equivalent to the search for a fully quantized system that is chaotic
 in a rigorous sense.  Defining chaos as \begin{itshape}deterministic 
randomness\end{itshape} may prove to be very useful in this effort.  (With 
this definition, a system is nonchaotic if the information contained in a 
system is logarithmically compressed by the algorithms use to compute it.  
That is, one avoids the use of Lyapunov exponents by comparing the size of 
information input to the size of information output.)\cite{ilg}

	As discussed earlier, the notion of Lyapunov exponents does not 
directly carry over to the quantum regime.  Even in the semiclassical or high 
quantum number regimes of quantum chaology, one still has problems defining 
this concept.  Nevertheless, it is desirable to have a notion of stability 
that can be used for these situations.  In classical systems, the Lyapnuov 
exponent is defined as the rate at which the largest eigenvalue of a 
trajectory grows.  When the Lyapunov exponent of a classical trajectory is 
positive, the associated motion exhibits exponential sensitivity.  In 
computing these exponents, one must consider how fast neighboring trajectories
 spread apart, which leads to difficulties as discussed above.  (Additionally,
 one no longer has trajectories in the traditional sense.)  One may thus 
define an \begin{itshape}instability exponent\end{itshape} $\chi$ related to 
the eigenvalues of trajectories near periodic orbits.  (They are computed in 
the semiclassical regime for so-called ``periodic orbit expansions'' and the 
Gutzwiller trace formula.)  These instability exponents are the needed 
generalization of Lyapunov exponents.  We remark that though we use these 
instability exponents in quantum chaology, there are fundamentally classical 
objects that are based on the concept of periodic orbits.  In order to use 
classical periodic orbits in quantum mechanics, one must calculate the action 
integral $S$ over one period and also consider its variation $\delta S$.  One 
uses the Bohr correspondence priniple to intepret the expression one obtains 
for $\delta S$.  Moreover, when a classically chaotic system is quantized, one
 obtains \begin{itshape}scars\end{itshape}, which describe the narrow linear 
regions with enhanced intensity that occur in an eigenstate's intensity 
pattern.\cite{gutz}

	Before we indulge ourselves in quantum chaology, let's consider some 
alternate terminology in the generalization of the notion of classical chaos. 
 In so doing, we introduce the notion of 'quantum billiard chaos,'\cite{qbc} 
which may also be described in terms of the nodal properties of wavefunctions.
  This notion of chaos corresponds to quantized chaology as it manifests in 
quantum billiards.  (As we discuss later, this can occur in billiards for 
which the Helmholtz equation is not globally separable.)  It is chaotic in the
 sense of boundedness, infinite recurrence, and the instability exponents just
 defined.  We earlier generalized the notion of chaos directly to the quantum 
regime.  In this alternate formulation, one retains the classical notion of 
chaos and instead generalizes one if its component conditions--the idea of 
exponential sensitivity.  Though this procedure is reasonable, we will instead
 partition quantum chaos into three categories as discussed above, because 
doing so allows us to discuss quantum chaos for a wider class of systems.

	The tools used in the study of quantum chaology include random matrix 
theory, level dynamics, and periodic orbit expansions.  The former comes into 
play in considering a system's (Hermitian) Hamiltonian matrix $H$.  One 
defines an ``uncorrelated'' probability density $p(H)$.  That is, separate 
blocks of $H$ are uncorrelated so that if $H$ is an $n \times n$ matrix, then
\begin{equation}
	p(H) = \prod_{i = 1}^n \prod_{j = i}^n p_{ij}(H_{ij}). 
\label{ind}
\end{equation}
For the special case in which $n = 2$, equation (\ref{ind}) becomes
\begin{equation}
	p(H) = p_{11}(H_{11})p_{12}(H_{12})p_{22}(H_{22}).
\end{equation}	
One can show that
\begin{equation}
	p(H) = C \exp\left( - A \text{ tr}(H^2)\right),
\end{equation}
where $A$ and $C$ are constants of integration.  Related to this is the Wigner
 distribution 
\begin{equation}
	P_W(x) = \frac{\pi}{2}x\exp\left(-\frac{\pi}{4}x^2\right).
\end{equation}
One expects such Wignerian statistics to hold for the spectra of complicated 
quantum systems with many degrees-of-freedom, in which the associated 
Hamiltonian is similarly complicated.  Wigner statistics $P_W$ are valid only 
if a system has integral spin and is invariant under an anti-unitary 
transformation such as time reversal.  (An anti-unitary transformation 
consists of the composition of a unitary transformation with complex 
conjugation.\cite{sakurai})  Among the appropriate complex quantum systems are
 atoms and atomic nuclei.  It has been shown that Wigner's prediction is 
consistent with experimentally obtained data describing the spacing of energy 
levels.  The hydrogen atom in a strong magnetic field is a ``simple'' system 
in which this behavior is observed.  The underlying chaotic features of the 
system cause certain statistical features of the quantum spectrum to obey 
predictions from random matrix theory.\cite{atomic}

	If each of the anti-unitary symmetries is broken, the nearest 
neighbor statistics are expected to be described by
\begin{equation}
	P_U(x) = \frac{32x^2}{\pi^2}\exp\left(-\frac{4x^2}{\pi}\right),
\end{equation}
where the subscript `$U$' stands for `unitary' since in the present case the 
system's Hamiltonian is invariant under all unitary transformations.  If the 
system has half-integral spin and retains anti-unitary symmetry, then all 
levels of the system are degenerate in the sense of Kramers.\cite{kramers}  
The nearest neighbor distribution of energy level spacings between degenerate 
pairs is given by
\begin{equation}
	P_S(x) = \frac{2^{18}x^4}{3^6\pi^3} \exp\left(-\frac{64x^2}{9\pi}
\right)
\end{equation}
where the subscript `$S$' stands for `symplectic'.  If all the anti-unitary 
symmetries are broken, we are again in the case $P_U$.
	
	The three cases $P_W$, $P_U$, and $P_S$ are characterized by the Dyson
 parameter $\beta$, which indicates the degree of level repulsion as $x 
\longrightarrow 0$.  In the three types of statistics above, $\beta$ takes the
 respective values 1, 2, and 4 for the probability densities $P_W$, $P_U$, and
 $P_S$.  The matrix elements of the Hamiltonian have a Gaussian distribution 
and for $\beta = 1, 2, 4$ are invariant respectively under orthogonal, 
unitary, and symplectic transformations.  The random matrix ensembles are 
given the respective names Gaussian orthogonal ensemble (GOE), Gaussian 
unitary ensemble (GUE), and Gaussian symplectic ensemble (GSE).  In 
addition to determining the appropriate level statistics, the symmetry class 
of a given Hamiltonian $H$ also influences physical characteristics such as a 
system's localization length (which determines how fast wavefunctions decay in
 a given basis).  Additionally, the zeroes of the Riemann $\zeta$-function 
correspond very closely with the GUE\cite{oldgutz}, which leads to the 
well-known connection between random matrix theory and this famous function.

	The concept of level dynamics is also useful in studying the 
quantum-mechanical analogs of classically chaotic systems.   If a quantum 
system depends on an external parameter, such as the strength of an externally 
applied magnetic field, its energy levels depend on that parameter.  One may 
examine the changes in these energy levels as that parameter is adjusted.  
These so-called \begin{itshape}level dynamics\end{itshape} provide useful 
information about a quantum-mechanical system's underlying chaotic structure. 
 One studies the eigenenergies $E_n$ of a Hermitian Hamiltonian
\begin{equation}
	H(\epsilon) = H_0 + \epsilon V \label{pert}
\end{equation}
as a function of the perturbation parameter $\epsilon$.  (The functions $H_0$ 
and $V$ are independent of $\epsilon$.)  One identifies the eigenenergies 
$E_n$ with fictitious particles.  To derive the equations of motion, one 
starts with the eigenvalue equation
\begin{equation}	
	H(\epsilon)|n(\epsilon)\rangle = E_n(\epsilon)|n(\epsilon)\rangle
\end{equation}
and defines the matrix elements
\begin{equation}
	V_{nm}(\epsilon) = \langle n(\epsilon)|V|m(\epsilon)\rangle.
\end{equation}
Assuming that $H$ is invariant under time reversal guarantees that eigenstates
 and matrix elements are real.  The evolution equations of the level dynamics 
are then 
\begin{gather}
	\dot{E}_n(\epsilon) = V_{nn}(\epsilon), \notag \\
	\dot{V}_{nn} = 2\sum_{m\neq n}\frac{V_{nm}^2}{E_n - E_m}, \notag \\
	\dot{V}_{nm} = \frac{V_{nm}(V_{nn} - V_{mm})}{E_m - E_n} + 
\sum_{l\neq n,m}V_{nl}v_{lm}\left(\frac{1}{E_n - E_l} + 
\frac{1}{E_m - E_l}\right), \label{level}
\end{gather}
which can also be derived from Hamilton's equations.  These equations can be 
solved once the initial conditions at $\epsilon = 0$ are known.  Note that the
 structure of equation (\ref{level}) is independent of the specific form of 
the Hamiltonian $H$.  Moreover, equation (\ref{level}) shows that the dynamics
 of energy levels of all Hamiltonian systems can be split into two parts, 
corresponding to equation (\ref{pert}).  (This separation of the Hamiltonian 
into an unperturbed Hamiltonian $H_0$ plus a small perturbation $\epsilon V$ 
is reminiscent of Melnikov theory.\cite{gucken,wiggins})  The characteristics 
of a given Hamiltonian are used only via the initial conditions.  Thus, if the 
initial conditions are unimportant for a sufficiently large $\epsilon \geq 
\hat{\epsilon}$, then one can use equilibrium statistical mechanics to compute
 the statistical properties of the system's energy levels.  The methods of 
level dynamics can be generalized to those of ``resonance 
dynamics.''\cite{atomic}  In equation (\ref{level}), the indices $n$ and $m$ 
range over the number of dimensions of the relevant Hilbert space 
$\mathcal{H}$.  These equations preserve the Liouville volume, the energy 
\begin{equation}
	E = \frac{1}{2}\sum_n V_{nn}^2 + \frac{1}{2}\sum_{n \neq m} 
\frac{|V_{nm}|^2}{(E_n - E_m)^2},
\end{equation}
and the total coupling strength
\begin{equation}
	Q = \sum_{n \neq m}|V_{nm}|^2,  
\end{equation}
which gives infinitely many first integrals (constants of motion), since 
increasing the coupling strength $\epsilon$ generates an orthogonal 
transformation in $\mathcal{H}$ whose invariants (including the traces of 
various operators) stay the same.  Moreover, the dynamical system 
(\ref{level}) is completely integrable, so that one must understand a system 
with infinitely degrees-of-freedom and infinitely many constants of motion.  
In order to do this, one could take the point of view of statistical 
mechanics.  Consider a stationary distribution
\begin{equation}
	P = \frac{1}{Z} e^{-\beta E - \gamma Q},
\end{equation}
where $Z$ is the partition function (a normalization constant).  The `inverse 
temperature' $\beta$ and `chemical potential' $\gamma$ are determined by 
prescribing mean values $\bar{E} \equiv \langle E \rangle$ and $\bar{Q} \equiv
 \langle Q \rangle$.  One integrates out the variables $V_{nm}$ to give a 
distribution of the eigenvalues $E_n$  One obtains a GOE probability 
distribution in the eigenvalues, which is a bit disturbing because this 
ensemble now seems to show up in a context sufficiently general that it may 
not be of much value to the characterization of quantum signatures of 
classical chaos.\cite{gutz}

	A third tool used in quantum chaology is the study of periodic orbit 
expansions.  In order to pursue this field, it was essential to develop 
semiclassical methods that worked in the quantum regime.  The first procedure 
to do this was the formalism of periodic orbit quantization.\cite{gutz}  The 
central result of this theory is the Gutzwiller ``trace formula.''  In the 
semiclassical approximation, only periodic orbits $\{p\}$ contribute in the 
evaluation of the level density
\begin{equation}
	\rho(E) \equiv \text{tr}\left[\delta(E - H)\right].
\end{equation}
One finds that the classical approximation $\rho_c(E)$ of the 
quantum-mechanical trace\cite{gutz} is 
\begin{equation}
	\rho_c(E) = \frac{1}{i\hbar}\sum_p\frac{T_0}{2 \sinh(\chi/2)} 
\exp\left[i\left(\frac{S}{\hbar} - l\frac{\pi}{2}\right)\right], \label{trace}
\end{equation}
where $E$ represents energy,
\begin{equation}
	T_0(E) \equiv \int \frac{dq_1}{|\dot{q}|}
\end{equation}
 is the primitive period, $\chi(E)$ is the instability exponent, $S(E) = \int 
pdq$ (integrated over the periodic orbit) is the action integral, and $l$ is 
the number of times the stable manifold is oriented in the local 
$p$-direction.  The superposition of these smooth classical functions yields 
an approximate spectrum for the quantum-mechanical energy levels.  We remark 
that the above formulation corresponds to a Feynman path integral approach, so
 that we are integrating in the variable $q \equiv (q_1,q_2,q_3)$ from some 
initial point $q'$ to some terminal point $q''.$  This formulation has been 
generalized, but by considering a Green's function (that is, a 
propagator\cite{sakurai,butkov}), even generalized versions of equation 
(\ref{trace}) reduce to a trace.  There are several other similar formulas in 
the study of quantized chaos, as discussed by Gutzwiller.\cite{gutz}  In some 
situations, for example, it is appropriate to have a hyperbolic cosine 
function rather than a hyperbolic sine function.  One can also study the 
Riemann-$\zeta$ function to gain insight into the trace formula.

	There are two ways in which the trace formula can be used to deal with 
quantum-mechanical problems.  It can be applied ``forwards'' to calculate the 
level density of a given quantum system based on purely classical input 
represented by the periods, actions, Lyapnuov exponents, and characteristic 
parameters of the associated classical periodic orbits.  It can also be 
applied ``backwards'' if the level density $\rho(E)$ is given.  In this usage,
 information about the periodic orbits is extracted from the level density by 
a generalized Fourier transform based on the trace formula (since the trace 
formula is in the form of an eigenfunction expansion).  The forward 
application of the trace formula is considered more difficult, although it has
 been used successfully on occasion.  Gutzwiller, for example, applied the 
trace formula to a system consisting of electrons with an asymmetric mass 
tensor moving in a Coulomb potential\cite{oldgutz}, an important problem in 
semiconductor physics.  The forward transform is considered difficult for 
three reasons:  The number of periodic orbits increases exponentially in a 
chaotic system, these orbits have to be computed numerically, and the trace 
formula has convergence issues that have to be circumvented with appropriate 
summation prescriptions.

	In the field of atomic physics, there are many systems whose classical
 counterparts are chaotic.  Perhaps the most famous one is the rotation of a 
diatomic molecule under the influence of externally applied 
microwaves.\cite{atomic,gutz}  A polar dimer molecule, such as CsI, is 
located between two plates of a capacitor, which is connected to a pulse 
generator that periodically charges and discharges the capacitor's two 
electrodes.  This process creates a time-varying, spatially homogeneous 
electric field, so that the molecule experiences a sequence of electric pulses
 that couple with its dipole moment.  This simple situation is a deterministic
 one, as we are assuming that there are no random fluctuations.  Consequently,
 given the initial state of the rotating diatomic molecule, one may compute 
its associated wavefunction for all time.

	The above technique captures the essential physical features of a 
rotating diatomic molecule, but it does not truly explain them.  In order to 
understand the dynamics of the present example, one approximates it by 
restricting its rotation to a single plane (rather than three-dimensional 
space) and by ignoring the motion of its center of mass.  In this 
approximation, one treats the rotating diatomic molecule as a 
\begin{itshape}kicked rotor\end{itshape}, an example that has become a 
paradigm of quantum chaology.  It has been studied in both a classical and 
quantum setting.\cite{atomic,gutz}  One may treat the classical kicked rotor 
as a pendulum (or a rotator) that is agitated at equal time intervals with an 
impulse that varies periodically as a function of the rotor's angular 
position.  The motion of the rotor is then uniform between kicks, each of 
which changes the system's angular momentum.  As the time interval between 
agitations becomes smaller, the behavior of the rotor approaches that of an 
ordinary forced pendulum.  This classical system behaves chaotically, and the 
effects of this behavior are evident in the behavior of its quantum-mechanical
 cousin.

	The study of quantum chaology and its applications remains an active 
area of research.\cite{atomic}  Theoretical research addresses its analytical 
structure as well as the convergence properties of Gutzwiller's trace formula 
and its derivatives.  A major breakthrough occurred in 1986 when Sir Michael 
Berry observed similarities between the trace formula and certain 
representations of Riemann's $\zeta$-function.  This function has henceforth 
served as a model for studying the analytical properties of semiclassical 
trace formulas.  There are more fundamental concerns, however, than the 
convergence of these formulas.  Indeed, it has been argued that there may 
exist a completely bounded chaotic dynamical system whose spectrum (which is 
both real and discrete) is identical to the imaginary parts of the 
(nontrivial) zeroes the Riemann $\zeta$-function.  The existence of this 
system would accomplish two important tasks.  As a mathematical problem, it 
would prove the Riemann conjecture that
\begin{equation}
	\zeta\left(\frac{1}{2} + iz\right) = 0
\end{equation}
in the region $|Im(z)| < 1/2$ only for $z$ with vanishing real part.  As a 
physical system, it would offer important insights regarding the analytical 
connection between classical chaos and quantum energy levels.  Research in 
applications of quantized chaos include attempts to use the semiclassical 
methods in this area as a mathematical tool for studying classically chaotic 
systems.  These methods are currently thought to be much more useful for 
interpreting quantum spectra and wavefunctions than they are for accurately 
predicting the spectra of classically chaotic systems.\cite{atomic}  

	There is active research in other areas of quantum chaology as well.  
For example, there have been several recent discoveries concerning the 
connection between quantized chaos and the three matrix ensembles discussed 
earlier.  Additionally, it is not completely understood why the statistics of 
random matrix ensembles are so accurate in describing how classical chaos 
induces universal fluctuations in energy levels.  Other topics of current 
interest include diffraction and refraction corrections in semiclassical 
procedures.  Some scientists are also studying whether the presence of chaos 
in a system can increase tunnelling.  In this context, a wave would tunnel 
between two islands in a chaotic sea.  It has been surmised that the presence 
of chaos would increase the tunnel splitting of energy levels by several 
orders of magnitude.  Lastly, quantum chaology in dimensions greater than two 
is virtually unexplored.  The methods in this subject are only expected to be 
accurate to order $\hbar^2$ independent of dimension, although this has not 
been shown rigorously.  The extension of quantum chaology to higher dimensions
 should nevertheless prove quite fruitful.  

\vspace{.1 in}

\begin{centering}
\subsection{Type II: Semiquantum Chaos}
\end{centering}

\vspace{-.1 in}

	Semiquantum chaos concerns systems with both classical and quantum 
subsystems.   It can arise, for example, in the form of the dynamic 
Born-Oppenheimer approximation,\cite{vibline} which shows up commonly in the 
study of mesoscopic and chemical physics.  This adiabatic approximation arises
 naturally in systems that may be expressed as the coupling of slow and fast 
subsystems.  In the Born-Oppenheimer scheme, chaos may occur in both the 
classical (slow) and quantum (fast) subsystems, although considered 
separately, neither of those regimes is necessarily chaotic.  The first step 
in the Born-Oppenheimer approximation is to quantize the fast (electronic) 
subsystem.  The second step is to quantize the nuclear (slow) subsystem.  If, 
however, the electronic energy levels are too close together, the 
Born-Oppenheimer approximation breaks down, as the electronic and nuclear 
subsystems are nonadiabatically coupled.  When this occurs, one treats the 
nuclear degrees-of-freedom as classical variables, thereby obtaining a 
semiquantal regime in which a classical system is coupled nonadiabatically to 
a quantum one.  It is this nonadiabatic coupling that produces semiquantum 
chaos in the resulting system, which is sometimes called a 
\begin{itshape}semiclassical quantization.\end{itshape}\cite{meyer,ezra}

	One may mathematically abstract numerous systems that exhibit 
semiquantum chaos as quantum billiards with vibrating 
boundaries.\cite{vibline,atomic,sazim,nec,rect,bif,ezra,ksu}  Such systems are
 not necessarily expressible \begin{itshape}precisely\end{itshape} as 
vibrating quantum billiards, but such billiards serve as a useful toy model in
 that that they capture many of the features of molecular systems.  It is in 
this abstract context that we discuss the notion of semiquantum chaos.

	Quantum billiards describe the motion of a point particle of mass $m_0$
undergoing perfectly elastic conditions in a bounded domain of mass $M \gg 
m_0$ in a potential $V$.  The particle's motion is described by the 
Schr\"odinger equation with Dirichlet boundary conditions.  One defines the 
``degree-of-vibration'' (\begin{itshape}dov\end{itshape}) of a billiard as the
 number of boundary dimensions that vary with time.  If the boundary is 
time-independent, the billiard is said to have zero 
\begin{itshape}dov\end{itshape}.  The one-dimensional vibrating billiard and 
the radially vibrating spherical billiard have a single
 \begin{itshape}dov\end{itshape}, and the rectanglular billiard with 
time-varying length and width has two \begin{itshape}dov\end{itshape}.  The 
\begin{itshape}dov\end{itshape} of a quantum billiard are its classical 
(``nuclear'') degrees-of-freedom. A zero \begin{itshape}dov\end{itshape} 
quantum billiard exhibits only integrable behavior if it is globally 
separable.\cite{nec}  (A quantum billiard is globally separable if the 
geometry of the billiard is one in which the Helmholtz equation is globally 
separable.)  Two simple ways in which global separability is violated are when
 a quantum billiard has a concave boundary component and when a billiard is 
geometrically composite, although it is believed that global separability may 
be violated in other ways (such as with a quantum billiard whose boundary is a
 quartic ellipse), by analogy with known non-integrable classical billiards.  
When global separability is violated in a quantum billiard, the observed 
behavior is ``chaotic'' in the sense of quantum chaology.  Billiards with 
concave boundary components, for example, share many of the properties of 
Anosov diffeomorphisms.\cite{arnold}  Composite quantum billiards such as the 
stadium billiard (whose boundary consists of two semi-circles joined by a pair
 of straight lines) have also been shown to exhibit chaotic 
behavior.\cite{mac,katok}  In globally separable, zero 
\begin{itshape}dov\end{itshape} quantum billiards, however, one expects to 
observe primarily quasiperiodic behavior, analogous to the Asteroids example 
discussed earlier.  (One must be careful with the term ``quasiperiodic'' in 
the quantum regime just as one is with the notion of chaos.  Esseentially, one
 is looking at the quantization of what was a quasiperiodic regime in the 
classical situation.)  The easiest example to visualize is that of a zero 
\begin{itshape}dov\end{itshape} rectangular quantum billiard.  Imagine that 
the boundary acts as a mirror, so that there are imaginary billiards adjoining
 the actual one on each of its four sides.  Modulo translation, one then 
recovers the same situation, as we described earlier when we discussed motion 
on a 2-torus.  If a wave hits a side of the billiard, the reflection in the 
mirror behaves just as the motion in the Asteroids example.  Under perfect 
reflection, the angle of incidence equals the angle of reflection, and so 
perfectly reflected trajectories have the same features as trajectories on a 
2-torus with respect to periodicity and quasiperiodicity.  Although a 
stationary, globally separable quantum billiard is necessarily integrable, we 
remark that the Toda lattice is an example of a dynamical system that is 
integrable but not separable.\cite{gutz,lich}

	Consider an $s$ \begin{itshape}dov\end{itshape} quantum billiard. 
 The total Hamiltonian of the system is given by   
\begin{equation}
	H(a_1, \cdots, a_s, P_1, \cdots, P_s) = K + \sum_{j = 1}^s 
\frac{P_j^2}{2M_j} + V,
\end{equation}
where $a_1, \cdots, a_s$ represent the time-varying boundary components and 
the kinetic energy (which corresponds to the quantum-mechanical Hamiltonian of
 the particle confined within the billiard) is given by  
\begin{equation}
	K = - \frac{\hbar^2}{2m} \nabla^2.
\end{equation}

	A two-term superposition of the $n$th and $q$th states (that 
is, a two-term Gal\"erkin projection\cite{infinite,gucken}) is given by
\begin{equation}
	\psi_{nq}(x,t) \equiv \alpha_n A_n(t)\psi_n(x,t) + 
\alpha_q A_q(t) \psi_q(x,t), \label{super}
\end{equation}
where the complex amplitudes $A_j(t)$ are time-dependent because the system 
has time-dependent boundary conditions.  Linear equations with such nonlinear 
boundary conditions thus lend themselves to analysis via Gal\"erkin methods 
just like nonlinear partial differential equations such as the Navier-Stokes 
and nonlinear Schr\"odinger equations.  The present problem is of a 
type known as a \begin{itshape}free-boundary problem\end{itshape}, in which 
one does not know a priori the shape of the domain.\cite{free}  It is 
well-posed by the specification of Dirichlet boundary conditions, an initial 
radius $a(0)$, and initial momentum $P(0)$, and initial amplitudes 
$A_j(0)$.

	For now, we specialize to the case of one 
\begin{itshape}dov\end{itshape}, in which only a single boundary dimension 
varies in time.  Porter and Liboff have shown\cite{nec} that a two-state 
superposition consisting of the $n$th and $q$th states exhibits chaotic 
behavior if and only if the quantum numbers corresponding to stationary 
dimensions of the billiard's boundary are the same in both states.  This 
result, which manifests in observed coupling behavior in the electronic states
 of polyatomic molecules\cite{ezra,bucky}, is easily extended to any 
finite-term superposition by considering the terms pairwise.  The result then 
states that there must exist one pair of states that satisfies the above 
condition.  In the vibrating spherical quantum billiard, for example, a pair 
of states has to have the same angular momentum quantum numbers $l$ and $m$ 
for chaotic behavior to occur.  (This result is proven using separability and 
orthogonality conditions of the Helmholz differential operator.\cite{nec})  
These conditions are not surprising, because the rotational symmetry of the 
system is invariant under radial vibrations.

	In an integrable two-term superposition state of a one 
\begin{itshape}dov\end{itshape} quantum billiard, the equations of motion are
\begin{equation}
	\dot{a} = \frac{P}{M},
\end{equation}
\begin{equation}
	\dot{P} = -\frac{\partial V}{\partial a} + \frac{\lambda}{a^3}, 
\label{inti}
\end{equation}
where $\lambda \equiv 2\left( \epsilon_1 |C_1|^2 + \epsilon_1 |C_2|^2 
\right)$, and $C_1$ and $C_2$ are constants such that $|C_1|^2 + |C_2|^2 = 1$.
  (The energy parameter $\lambda$ is necessarily positive because $\epsilon_i 
> 0$ and the $|C_i|^2$ correspond to probabilities.)  A special case of this 
configuration is obtained by considering a single eigenstate.

	Equation (\ref{inti}) has been studied numerically in the context of 
the bifurcations that can occur when one considers different potentials 
$V$.\cite{bif}  In particular, one observes only saddle-center bifurcations 
(and generalizations thereof).  Either all the equilibria are centers or--for 
sufficiently small energies $\lambda$--some of them are centers and others are
 saddle points (depending on the form of the potential).  Saddle connections 
in this system have been studied to some extent using continuation methods, 
although there is room for quite a bit more research in this area.

	In the chaotic case, the equations of motion take the form
\begin{equation}
	\dot{x} = -\frac{\omega_0 y}{a^2} - \frac{2 \mu P z}{Ma}, 
\label{xdot}
\end{equation}
\begin{equation}
	\dot{y} = \frac{\omega_0 x}{a^2},
\end{equation}
\begin{equation}
	\dot{z} = \frac{2 \mu P x}{Ma},
\end{equation}
\begin{equation}
	\dot{a} = \frac{P}{M}, 
\end{equation}
and
\begin{equation}
	\dot{P} = -\frac{\partial V}{\partial a} + \frac{2[\epsilon_+ + 
\epsilon_-(z - \mu x)]}{a^3}, \label{Pdot}
\end{equation}
where $x$, $y$, and $z$ are Bloch variables\cite{bloch}, $a$ represents a 
displacement, $P$ is its conjugate momentum, $M$ is the mass of the billiard, 
$m_0 \ll M$ is the mass of the confined particle, $\mu > 0$ is the coupling 
coefficient between the two eigenstates, $V \equiv V(a)$ is the potential in 
which the billiard resides, $\omega_0 \equiv (\epsilon_2 - \epsilon_1)/\hbar$, 
$\epsilon_{\pm} \equiv (\epsilon_2 \pm \epsilon_1)/2$, and 
$\epsilon_1$ and $\epsilon_2$ (where $\epsilon_2 \geq \epsilon_1$) are the 
energies of the two eigenstates.\cite{bif}  The bifurcation structure of this 
system of equations is a generalization of that observed in the integrable 
case in the sense that only generalized saddle-center bifurcations can occur. 
 As before, pairs of stable and unstable directions bifurcate to the center 
manifold as the energy of the system is increased.  Additionally, there is 
evidence of saddle connections for this chaotic case.  

\begin{figure}[htb] 
	\begin{centering}
		\leavevmode
		\includegraphics[width = 2.5 in, height = 3 in]{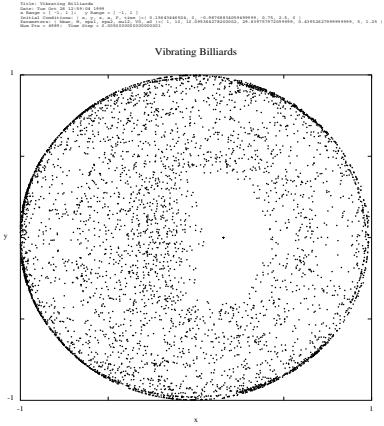}

\vspace{-.5 in}

		\caption{An example of chaotic behavior in the Bloch variables
 in a vibrating quantum billiard.} \label{fig4}
	\end{centering}
\end{figure}

	For the one-dimensional vibrating quantum billiard, Blumel and his 
co-authors\cite{vibline,atomic} considered a two-term superposition state that
 they computed to have a coupling coefficient $\mu = 3/4$.  They computed 
Liapunov exponents to show expontial divergence (and hence chaos) in this 
case.  However, for given parameter values and initial conditions, one can 
tell whether the configuration is chaotic simply by examining the associated 
Poincar\'e maps.  This, of course, does not \begin{itshape}prove\end{itshape} 
rigorously that the configuration is chaotic.  Nevertheless, it corresponds to
 the canonical application of KAM theory to engineering and the physical 
sciences, so we consider this a sufficient demonstration of chaotic behavior 
in the present context.  Figures \ref{fig2} and \ref{fig3} display chaotic 
behavior in the classical variables, and Figure \ref{fig4} shows chaos in the 
quantum-mechanical Bloch variables.  Classical Hamiltonian chaos in the 
positions and momenta lead to quantum-mechanical wave chaos in the normal 
modes, whereas chaos in the Bloch variables corresponds to chaos in the 
quantum probabilities.\cite{nec}

	The single \begin{itshape}dov\end{itshape} quantum billiard that has 
been analyzed most extensively is the vibrating sphere\cite{sazim}, 
although the one-dimensional vibrating billiard was studied 
earlier.\cite{vibline,atomic}  One finds that the coupling coefficient $\mu$ 
depends on the geometry of the system (as well as what quantum states one 
considers), but that the general behavior of the system is typified by the 
fact that only a single boundary variable is time-dependent.  Even though the 
vibrations of a sphere occur in $\mathbb{R}^3$, only the radial dimension of 
the billiard actually depends on time.  Other single 
\begin{itshape}dov\end{itshape} quantum billiards include the radially 
vibrating cylinder, the longitudinally vibrating billiard, and the rectanglar 
billiard in which either the length or width (but not both) depends on time.  

	One can generalize the above notions to quantum billiards with two or 
more \begin{itshape}dov\end{itshape}.  Two examples of such quantum billiards 
are the rectangular billiard in which both length and width are permitted to 
vibrate\cite{rect} and the cylindrical billiard with both radial and 
longitudinal vibrations.  The latter may be useful as a model for carbon 
nanotubes or quantum wires.  The theorems cited above can be generalized, 
although these generalizations require further study.

	Before we proceed to the third type of quantum chaos, it is important 
to discuss what is meant when we say that vibrating billiards exhibit quantum 
chaotic behavior when the coupling coefficient $\mu$ is nonzero.  What exactly
 constitutes such behavior and why should one care about it?  We formulated 
the chaotic configuration of single \begin{itshape}dov\end{itshape} quantum 
billiards as a five-dimensional dynamical system: the displacement $a$ and 
conjugate momentum $P$ are classical variables, and the Bloch variables $x$, 
$y$, and $z$ are quantum-mechanical variables--as they are obtained from the 
probabilities $|A_n|^2$ and $|A_q|^2$.  Another way to formulate this system 
is to use action-angle variables, which gives one classical degree of freedom 
coupled to a single quantum-mechanical one.\cite{ezra}  In each 
interpretation, we treat the dynamical equations as a classical subsystem 
coupled to a quantum one.  The chaotic behavior in the $(a,P)$-plane 
represents wave chaos, as the position $a$ is an argument of the individual 
wavefunctions $\psi_n$ and $\psi_q$.  Additionally, chaos on the Bloch sphere 
(equivalently, in the quantum-mechanical action and angle) represents chaos in
 the quantum probabilities.  Therefore, both manifestations of the observed 
chaotic behavior have interpretations that go beyond classical Hamiltonian 
chaos.  Moreover, the observed chaotic amplitudes along with chaotic waves 
give us a chaotic superposition of chaotic normal modes.  This is a hallmark 
of semiquantum chaos.  

	Vibrating quantum billiards, though an idealized mathematical model, 
bear import as simple manifestations of semiquantum chaotic behavior.  They 
are important for other--more practical--reasons as well.  They may be used to
 describe nonadiabatic coupling in polyatomic molecules.  Such behavior is 
relevant, for example, in the study of Jahn-Teller systems.\cite{ezra,ezra2}  
In the present context, chaotic configurations of a one 
\begin{itshape}dov\end{itshape} quantum billiard are analogous to a diatomic
molecule with two electronic states (of the same symmetry) coupled
nonadiabatically by the single internuclear vibrational coordinate.\cite{ezra}
  The former gives the quantum degrees-of-freedom, while the latter produces 
the classical ones.  (This situation is easily generalized to ones with three 
or more electronic states.  Additionally, the radially vibrating spherical 
quantum billiard captures features of particle behavior in the 
nucleus\cite{wong} and as a simplistic representation of the quantum dot 
nanostructure.\cite{qdot}  The vibrating cylindrical billiard may be used as a
 model for the quantum wire, another microdevice component.\cite{qwire}  More 
importantly, it may also prove useful as a model of carbon nanotubes.  
Additionally, other geometries of vibrating quantum billiards may have similar
 applications in mesoscopic physics.  That is, they may be used as models of 
various chemical nanostructures, as they describe the nanomechanical 
(electronic-vibronic) coupling that can occur in such devices.\cite{bucky}  
Moreover, vibrating quantum billiards generalize Enrico Fermi's bouncing ball 
model of cosmic ray acceleration\cite{fermi,lich}.  Finally, when some 
compounds are placed into liquids, one obtains solvated electrons that may be 
described as oscillating billiard systems.\cite{solve1,solve2}  With such a 
wide array of possible application, vibrating quantum billiards are a very 
versatile model.  They provide a simple illustration of semiquantum chaos, 
they generalize a toy model of Enrico Fermi, and they have already been shown 
to be relevant to areas of nuclear, chemical, and mesoscopic physics.

	There remains much to be studied about semiquantum chaos in vibrating 
billiard systems.  For example, one may analyze Gal\"erkin projections with 
more than two terms (in which the Bloch sphere is generalized) and quantum 
billiards with two and three \begin{itshape}dov\end{itshape}.  A quantum 
billiard with three \begin{itshape}dov\end{itshape} (such as a rectangular 
prism billiard with vibrating length, width, and depth) may also exhibit 
Arnold diffusion and cross-resonance diffusion, because the Hamiltonian 
(classical) subsystem has a number of degrees-of-freedom equal to the number 
of \begin{itshape}dov\end{itshape}s of the billiard.  These problems may also 
be studied using action-angle coordinates, in which the number of 
degrees-of-freedom of the system is more readily apparent.  Future work also 
includes the study of other geometries, such as the two 
\begin{itshape}dov\end{itshape} cylindrical quantum billiard and billiards 
with concentric geometries (such as spheres or disks) whose inner and outer 
radi both oscillate.  Note, in particular, that this latter example removes 
the assumption of convexity and may also lead to a generalization of the 
notion of \begin{itshape}dov\end{itshape}.  Another effect to incorporate is 
that of rebound from the collisions of particles with the billiard's boundary.
  Additional work to be done includes further studies of bifurcations in 
vibrating quantum billiards as well as an analysis of coupled vibrating 
quantum billiards, which is important because quantum dots are often coupled 
in arrays of various geometries in laboratory settings.  It may also be 
fruitful to extend the analysis of vibrating quantum billiards to a 
relativistic setting as well as to studying multiple-particle vibrating 
quantum billiards.  Finally, the Gal\"erkin method discussed briefly in the 
present paper may also be useful for analyzing nonlinear Schr\"odinger 
equations, because linear partial differential equations with nonlinear 
boundary conditions are similar in several respects to nonlinear partial 
differential equations.

	Semiquantum chaos may be considered in other settings as well.  The 
fact that such systems exhibit exponential sensitivity in their quantum 
subsystems (represented by the fact that the Bloch variables behave 
chaotically) is a hallmark of the traditional notion of chaos in a semiquantal
 setting.\cite{atomic,nec}  We remark that quantum chaos is a type of `wave 
chaos.'  This generalization is nominal in terms of `type I' quantum chaos, as
 it make little difference whether a system is quantum-mechanical or classical
 when one is studying spectra or wave manifestations of ray dynamics in 
classical wave systems in areas such as acoustics, optics, and 
electrodynamics.  However, when applying this generalization to the 
semiquantal regime, it is important to note that for classical wave systems, 
having a classical boundary is no longer an approximation.  This shows that 
`type II' wave chaos exists in nature.  Such waves have been studied in 
classical electrodynamics.  

	It would be fruitful to apply the methods that have been used in the 
study of semiquantum chaos to nonlinear Schr\"odinger equations, which can be 
used to describe optical waves in certain media as well as superfluid 
hydrodynamics.\cite{nls}  Recall, however, that in quantum mechanics, the 
classical boundaries used in the present analysis are in truth an 
approximation.  The nuclear degrees-of-freedom (represented by the oscillating
 components of the billiard boundary) may be quantized, resulting in a purely 
quantum (though higher-dimensional) system.  The effect of such a quantization
 on semiquantum chaotic systems has not been completely resolved, although 
every previous attempt has produced an example of quantized chaos.  In 
particular, quantizing the walls of a vibrating quantum billiard leads to a 
higher-dimensional system that properly falls under the heading of quantized 
chaos.

\begin{centering}
\subsection{Type III: True Quantum Chaos}
\end{centering}

	Bounded, fully quantized systems that exhibit exponential sensitivity 
and infinite recurrence are genuinely quantum chaotic.  (Semiquantum chaos 
describes exponential sensitivity in the semiquantal regime; systems in this 
regime consist of classical subsystems coupled with quantum mechanical ones, 
so they are not fully quantized.)  Quantum chaology describes quantum 
signatures of classically chaotic systems.  This regime is fully quantal, but 
the ``chaos'' observed cannot exhibit exponential sensitivity, as we discussed
 earlier.  Indeed, the existence of systems that are quantum chaotic in the 
above, stronger sense remains an open question.  From our previous discussion,
 we note that the energy spectrum of such a system cannot be fully discrete.  
Otherwise, such a system could not display exponential sensitivity.

	The following example has been proposed as a 
\begin{itshape}possibility\end{itshape} of such a quantum chaotic system.  
Most scientists do \begin{itshape}not\end{itshape} consider it an example of 
such, although it is certainly a very interesting system.  Consider a spin 1/2
 particle passing through a chain of two different magnets (types $A$ and 
$B$), sequenced according to the following recursion formula:
\begin{equation}
	M_{n+1} = M_n \circ M_{n - 1}, \hspace{.2 in} M_0 = A, M_1 = B,
\end{equation}	
where the symbol $\circ$ denotes the operation of appending one chain of 
magnets to another.  For example, $M_2 = M_1 \circ M_0 = BA$, $M_3 = M_2 \circ
 M_1 = BAB$, and $M_4 = M_3 \circ M_2 = BABBA$.

	In general, the chain of magnets (which, in principle, can be 
constructed in laboratory settings) induces spin precession.  The propagator 
$U$ of a spin 1/2 particle satisfies the recursion relation
\begin{equation}
	\hat{U}_{n+1} = \hat{U}_n\hat{U}_{n-1}
\end{equation}
and can be parametrized by
\begin{equation}
	\hat{U}_n = e^{-i\alpha_n\sigma_z} e^{-i\beta_n\sigma_y} 
e^{-i\gamma_n\sigma_z},	
\end{equation}
where $\sigma_x$, $\sigma_y$, and $\sigma_z$ are the Pauli spin 
matrices.\cite{sakurai}

	We follow Bl\"umel and Reinhardt and consider the special case in 
which the magnetic field of each of the magnets is aligned along the $y$-axis.
  It follows that the propagator $\hat{U}_n$ represents a rotation by angle 
$\beta_n$.  This leads to to following recurrence relation for $\beta$:
\begin{equation}
	\beta_{n+1} = \beta_n + \beta_{n - 1} \hspace{.2 in} 
(\text{mod } 2\pi).
\end{equation}
For the initial conditions $\beta_0 = \beta_1 = 1$, we recover the Fibonacci 
sequence.\cite{powerlaw}  Defining $b_n \equiv \beta_n/2\pi$, one obtains the 
above recursion relation mod 1, which can be written as the map $Q : 
\vec{w}_n \mapsto \vec{w}_{n+1} =$ 
\begin{equation}
	\begin{pmatrix}
		1 & 1 \\ 1 & 0
	\end{pmatrix}
   	\hspace{.2 in} (\text{mod }1), \label{map}
\end{equation}
where the vector $\vec{w} \equiv (b_n,b_{n-1})$.  The map (\ref{map}) is very 
similar to the Anosov map\cite{arnold} $C: \vec{w}_n \mapsto \vec{w}_{n+1} =$ 
\begin{equation}
	\begin{pmatrix}
		1 & 1 \\ 1 & 2
	\end{pmatrix}, \label{anosov}
\end{equation}
whose chaotic properties are well-known.  The map $Q$ shares many of the 
properties of the Anosov diffeomorphism.  For example, it possesses a 
stretching direction $\vec{v}_1^{(Q)} = (1,g)$ with a corresponding eigenvalue
 $e_1^{(Q)} = \bar{g} > 1$ that has a positive Lyapunov exponent.  The mapping
 $Q$ thus exhibits exponential sensitivity and chaotic behavior just like the 
Anosov diffeomorphism.  The quantum dynamics of spin 1/2 particles in the 
given magnetic chain are thus argued by Bl\"umel and Reinhardt to be truly 
chaotic.\cite{atomic}  The sequence of rotation angles $\beta_n 
(\text{mod }2\pi)$ is consequently also chaotic.  

	If the spin 1/2 particles are prepared in a pure spin state polarized 
in the $+z$ direction, the corresponding occupation probability in the 
$+|z\rangle$ state after the $n$th section is given by $\cos^2(\beta_n)$, so 
the population in the $+|z\rangle$ state must be chaotic as well.  Measuring 
this occupation probability provides an experimental test for the occurrence 
of quantum chaos in the present system.

	Additional issues are involved, which raises doubt as to whether one 
should consider this system a genuinely quantum chaotic one.  For a given $n$,
 the magnetic chain is not chaotic because of the unitarity of quantum 
mechanics, as there is no exponential instability for fixed $n$.  Instead, 
chaos occurs as a function of the discrete variable $n$.  Moreover, one must 
consider the length of the apparatus required to observe the quantum chaotic 
behavior described above.  It is well-known that the Fibonacci sequence 
diverges exponentially, so the number of magnets increases exponentially with 
$n$.  Therefore, the action of this chain of magnets is equivalent to free 
motion of a particle on a ring whose position is measured at the end of 
exponentially growing time intervals.  This provides an alternate means of 
understanding this example.  Because of the exponential increase in the length
 of the magnet chain, the ``physical flight time'' it takes for particles to 
go through the actual apparatus also grows exponentially in $n$.  However, if 
the magnets are exponentially close in $n$, then the ``natural flight time'' 
grows linearly, and the magnet chain would consequently also be chaotic with 
respect to this temporal variable.  Tranforming to the rest frame of the 
moving beam particles, the quantum-mechanical description of a spin 1/2 
particle traversing the chain of magnets is equivalent to the quantum 
description of a stationary spin 1/2 particle perturbed by a sequence of 
external field pulses.

	Note finally that the dynamics of the case with aligned magnetic 
fields is the only one that has been investigated thus far.  Additionally, 
this system of Fibonacci magnets is considered to be a genuinely quantum 
chaotic by only a handful of scientists.  Other candidate systems have been 
proposed\cite{chir}, but the existence of chaotic behavior (in the traditional
 sense) in fully quantized systems remains an open question.  Nevertheless, 
there are some clues concerning where to look.  In order to have a chance at 
representing the Holy Grail of quantum chaos, a system must be (spatially) 
bounded, finite-particle, undriven, and fully quantum with a spectrum that is 
not discrete.\cite{ilg}

\begin{centering}
\section{Conclusion}
\end{centering}

\vspace{-.1 in}

	The fields of nonlinear dynamics and quantum mechanics have both 
achieved their share of attention in popular culture.  Each has been a part of
 a scientific revolution.  For example, the notion of quantum mechanics 
brought an important probabilistic interpretation to science, and the advent 
of dynamical systems theory showed that determinism did not imply solvability.
  The study of quantum chaos, which has been increasingly scrutinized in 
recent years, is an effort to marry these two subjects.  In the present paper,
 we discussed the historical evolution and some principle ideas of both 
subjects.  We then divided quantum chaos into three behavioral subclasses and 
discussed several examples, methods, and results in each of these areas.

\vspace{-.1 in}

\begin{centering}
\section{Acknowledgements}
\end{centering}

	I would like to acknowledge Richard Liboff for advising me on my 
thesis, of which this paper will ultimately be the introduction.  He also 
suggested several improvements to an earlier draft of the present paper.  
Class notes from a stability and bifurcation course taught by Paul Steen were 
very helpful in the preparation of my discussion of those subjects in this 
paper.  Greg Ezra corrected several important mistakes and oversights in early
 drafts of this manuscript, and he also gave me several excellent ideas.  
Bruno Eckhardt also caught some errors in an earlier version of this 
manuscript.  Finally, I would like to thank Catherine Sulem for useful 
discussions concerning this project.

\begin{centering}
\bibliographystyle{plain}
\bibliography{ref}

\begin{thebibliography}{10}

\bibitem{mta}
Ralph Abraham, Jerrold~E. Marsden, and Tudor Ratiu.
\newblock {\em Manifolds, Tensor Analysis, and Applications}.
\newblock Number~75 in Applied Mathematical Sciences. Springer-Verlag, New
  York, NY, 2nd edition, 1988.

\bibitem{bloch}
L.~Allen and J.~H. Eberly.
\newblock {\em Optical Resonance and Two-Level Atoms}.
\newblock Dover Publications, Inc., New York, NY, 1987.

\bibitem{arnold}
Vladimir~I. Arnold.
\newblock {\em Geometrical Methods in the Theory of Ordinary Differential
  Equations}.
\newblock Number 250 in A Series of Comprehensive Studies in Mathematics.
  Springer-Verlag, New York, NY, 2nd edition, 1988.

\bibitem{fermi}
R.~Badrinarayanan and J.~V. Jos\'e.
\newblock Spectral properties of a {F}ermi accelerating disk.
\newblock {\em Physica D}, 83:1--29, 1995.

\bibitem{vibline}
R.~Bl\"umel and B.~Esser.
\newblock Quantum chaos in the {B}orn-{O}ppenheimer approximation.
\newblock {\em Physical Review Letters}, 72(23):3658--3661, 1994.

\bibitem{atomic}
R.~Bl\"umel and W.~P. Reinhardt.
\newblock {\em Chaos in Atomic Physics}.
\newblock Cambridge University Press, Cambridge, England, 1997.

\bibitem{butkov}
Eugene Butkov.
\newblock {\em Mathematical Physics}.
\newblock Addison-Wesley Publishing Company, Reading, MA, 1968.

\bibitem{chir}
B.~V. Chirikov, F.~M. Izrailev, and D.~L. Shepelyanksy.
\newblock Quantum chaos - localization vs ergodicity.
\newblock {\em Physica D}, 33:77--88, October-November 1988.

\bibitem{wall}
Doron Cohen.
\newblock Chaos and energy spreading for time-dependent hamiltonians, and the
  various regimes in the theory of quantum dissipation.
\newblock {\em Annals of Physics}, 283:175--231, 2000.

\bibitem{devaney}
Robert~L. Devaney.
\newblock {\em An Introduction to Chaotic Dynamical Systems}.
\newblock Addison-Wesley, Redwood City, CA, 2nd edition, 1989.

\bibitem{diacu}
Florin Diacu and Philip Holmes.
\newblock {\em Celestial Encounters: The Origins of Chaos and Stability}.
\newblock Princeton University Press, Princeton, NJ, 1996.

\bibitem{feynman1}
Richard~P. Feynman, Robert~B. Leighton, and Matthew Sands.
\newblock {\em The {F}eynman Lectures on Physics}, volume~I.
\newblock Addison-Wesley Publishing Company, Reading, MA, 1964.

\bibitem{feynman3}
Richard~P. Feynman, Robert~B. Leighton, and Matthew Sands.
\newblock {\em The {F}eynman Lectures on Physics}, volume III.
\newblock Addison-Wesley Publishing Company, Reading, MA, 1964.

\bibitem{ilg}
Joseph Ford and Matthias Ilg.
\newblock Eigenfunctions, eigenvalues, and time evolution of finite, bounded,
  undriven, quantum systems are not chaotic.
\newblock {\em Physical Review A}, 45(9):6165--6173, May 1992.

\bibitem{free}
Avner Friedman.
\newblock Free boundary problems in science and technology.
\newblock {\em Notices of the American Mathematical Society}, 47(8):854--861,
  September 2000.

\bibitem{gleick}
James Gleick.
\newblock {\em Chaos: Making a New Science}.
\newblock Penguin USA, New York, NY, 1988.

\bibitem{goldstein}
Herbert Goldstein.
\newblock {\em Classical Mechanics}.
\newblock Addison-Wesley Publishing Company, Reading, MA, 2nd edition, 1980.

\bibitem{gucken}
John Guckenheimer and Philip Holmes.
\newblock {\em Nonlinear Oscillations, Dynamical Systems, and Bifurcations of
  Vector Fields}.
\newblock Number~42 in Applied Mathematical Sciences. Springer-Verlag, New
  York, NY, 1983.

\bibitem{oldgutz}
Martin Gutzwiller.
\newblock Periodic orbits and classical quantization conditions.
\newblock {\em Journal of Mathematical Physics}, 12:343--358, 1971.

\bibitem{gutz}
Martin~C. Gutzwiller.
\newblock {\em Chaos in Classical and Quantum Mechanics}.
\newblock Number~1 in Interdisciplinary Applied Mathematics. Springer-Verlag,
  New York, NY, 1990.

\bibitem{haake}
Fritz Haake.
\newblock {\em Quantum Signatures of Chaos}.
\newblock Springer Series in Synergetics. Springer-Verlag, Berlin, Germany, 2nd
  edition, 2001.

\bibitem{katok}
Anatole Katok and Boris Hasselblatt.
\newblock {\em Introduction to the Modern Theory of Dynamical Systems}.
\newblock Cambridge University Press, New York, NY, 1995.

\bibitem{kramers}
H.~A. Kramers.
\newblock \"uber das modell des heliumatoms.
\newblock {\em Z. Phys.}, 13:312--341, 1923.

\bibitem{liboff}
Richard~L. Liboff.
\newblock {\em Introductory Quantum Mechanics}.
\newblock Addison-Wesley, San Francisco, CA, 3rd edition, 1998.

\bibitem{kinetic}
Richard~L. Liboff.
\newblock {\em Kinetic Theory: Classical, Quantum, and Relativistic
  Descriptions}.
\newblock Wiley, New York, NY, 2nd edition, 1998.

\bibitem{qbc}
Richard~L. Liboff.
\newblock Quantum billiard chaos.
\newblock {\em Physics Letters}, A269:230--233, 2000.

\bibitem{sazim}
Richard~L. Liboff and Mason~A. Porter.
\newblock Quantum chaos for the radially vibrating spherical billiard.
\newblock {\em Chaos}, 10(2):366--370, 2000.

\bibitem{lich}
Allan~J. Lichtenberg and M.~A. Lieberman.
\newblock {\em Regular and Chaotic Dynamics}.
\newblock Number~38 in Applied Mathematical Sciences. Springer-Verlag, New
  York, NY, 2nd edition, 1992.

\bibitem{qdot}
J~Lucan.
\newblock {\em Quantum Dots}.
\newblock Springer, New York, NY, 1998.

\bibitem{mac}
S.~W. MacDonald and A.~N. Kaufman.
\newblock Wave chaos in the stadium: Statistical properties of short-wave
  solutions of the {H}elmholtz equation.
\newblock {\em Physical Review A}, 37:3067, 1988.

\bibitem{analysis}
Jerrold~E. Marsden and Michael~J. Hoffman.
\newblock {\em Elementary Classical Analysis}.
\newblock W. H. Freeman and Company, New York, NY, 2nd edition, 1993.

\bibitem{ms}
Jerrold~E. Marsden and Tudor~S. Ratiu.
\newblock {\em Introduction to Mechanics and Symmetry}.
\newblock Number~17 in Texts in Applied Mathematics. Springer-Verlag, New York,
  NY, second edition, 1999.

\bibitem{merz}
Eugen Merzbacher.
\newblock {\em Quantum Mechanics}.
\newblock John Wiley and Sons, Inc., New York, NY, 3rd edition, 1998.

\bibitem{meyer}
Hans-Dieter Meyer and William~H. Miller.
\newblock A classical analog for electronic degrees of freedom in nonadiabatic
  collision processes.
\newblock {\em Journal of Chemical Physics}, 70(7):3214--3223, April 1979.

\bibitem{break}
F.~L. Moore, J.~C. Robinson, C.~F. Bharucha, Bala Sundaram, and M.~G. Raizen.
\newblock Atom optics realization of the quantum $\delta$-kicked rotor.
\newblock {\em Physical Review Letters}, 75(25):4598--4601, December 1995.

\bibitem{bucky}
Hongkun Park, Jiwoong Park, Andrew~K. Lim, Erik~H. Anderson, A.~Paul
  Alivisatos, and Paul~L. McEuen.
\newblock Nanomechanical oscillations in a single ${C}_{60}$ transistor.
\newblock {\em Nature}, 407:57--60, September 2000.

\bibitem{celest}
Mason~A. Porter.
\newblock A historical approach to dynamical systems through celestial
  mechanics.
\newblock Unpublished, January 2000.

\bibitem{bif}
Mason~A. Porter and Richard~L. Liboff.
\newblock Bifurcations in one degree-of-vibration quantum billiards.
\newblock {\em International Journal of Bifurcation and Chaos}, 11(4):903--911,
  April 2001.

\bibitem{ksu}
Mason~A. Porter and Richard~L. Liboff.
\newblock The radially vibrating spherical quantum billiard.
\newblock {\em Discrete and Continuous Dynamical Systems}, pages 310--318,
  2001.
\newblock Proceedings of the International Conference on Dynamical Systems and
  Differential Equations: Georgia, May 18-21, 2000.

\bibitem{rect}
Mason~A. Porter and Richard~L. Liboff.
\newblock Quantum chaos for the vibrating rectangular billiard.
\newblock {\em International Journal of Bifurcation and Chaos}, To appear
  September, 2001.

\bibitem{nec}
Mason~A. Porter and Richard~L. Liboff.
\newblock Vibrating quantum billiards on {R}iemannian manifolds.
\newblock {\em International Journal of Bifurcation and Chaos}, To appear
  September, 2001.

\bibitem{sakurai}
Jun~John Sakurai.
\newblock {\em Modern Quantum Mechanics}.
\newblock Addison-Wesley Publishing Company, Reading, MA, {R}evised edition,
  1994.

\bibitem{powerlaw}
Manfred Schroeder.
\newblock {\em Fractals, Chaos, Power Laws: Minutes from an Infinite Paradise}.
\newblock W. H. Freeman and Company, New York, NY, 1991.

\bibitem{simmons}
George~F. Simmons.
\newblock {\em Differential Equations with Applications and Historical Notes}.
\newblock McGraw-Hill, Inc., New York, NY, 2nd edition, 1991.

\bibitem{solve1}
B.~Space and D.~F. Coker.
\newblock Nonadiabatic dynamics of excited excess electrons in simple fluids.
\newblock {\em Journal of Chemical Physics}, 94(3):1976--1984, February 1991.

\bibitem{solve2}
B.~Space and D.~F. Coker.
\newblock Dynamics of trapping and localization of excess electrons in simple
  fluids.
\newblock {\em Journal of Chemical Physics}, 96(1):652--663, January 1992.

\bibitem{strogatz}
Steven~H. Strogatz.
\newblock {\em Nonlinear Dynamics and Chaos}.
\newblock Addison-Wesley, Reading, MA, 1994.

\bibitem{nls}
Catherine Sulem and Pierre-Louis Sulem.
\newblock {\em The Nonlinear {S}chr\"odinger Equation: Self-Focusing and Wave
  Collapse}.
\newblock Number 139 in Applied Mathematical Sciences. Springer-Verlag, New
  York, NY, 1999.

\bibitem{infinite}
Roger Temam.
\newblock {\em Infinite-Dimensional Dynamical Systems in Mechanics and
  Physics}.
\newblock Number~68 in Applied Mathematical Sciences. Springer-Verlag, New
  York, NY, 2nd edition, 1997.

\bibitem{ezra}
Rober~L. Whetten, Gregory~S. Ezra, and Edward~R. Grant.
\newblock Molecular dynamics beyond the adiabatic approximation: New
  experiments and theory.
\newblock {\em Annual Reviews of Physical Chemistry}, 36:277--320, 1986.

\bibitem{wiggins}
Stephen Wiggins.
\newblock {\em Introduction to Applied Nonlinear Dynamical Systems and Chaos}.
\newblock Number~2 in Texts in Applied Mathematics. Springer-Verlag, New York,
  NY, 1990.

\bibitem{wong}
S~Wong.
\newblock {\em Introductory Nuclear Physics}.
\newblock Prentice Hall, Englewood Cliffs, NJ, 1990.

\bibitem{qwire}
H~Zaren, K~Vahala, and A~Yariv.
\newblock Gain spectra of quantum wires with inhomogeneous broadening.
\newblock {\em IEEE Journal of Quantum Electronics}, 25:705, 1989.

\bibitem{ezra2}
Josef~W. Zwanziger, Edward~R. Grant, and Gregory~S. Ezra.
\newblock Semiclassical quantization of a classical analog for the
  {J}ahn-{T}eller {E} $\times$ e system.
\newblock {\em Journal of Chemical Physics}, 85(4):2089--2098, August 1986.

\end{thebibliography}
\end{centering}

\begin{centering}
\subsection*{Figure Captions}
\end{centering}

Figure 1: A separatrix that occurs in an integrable configuration of a 
vibrating quantum billiard in a double-well potential.  Trajectories inside 
the separatrix behave qualitatively differently from those outside the 
separatrix.

\vspace{.1 in}

Figure 2: An example of hard Hamiltonian chaos.

\vspace{.1 in}

Figure 3: An example of soft Hamiltonian chaos.

\vspace{.1 in}

Figure 4: An example of chaotic behavior in the Bloch variables in a vibrating
 quantum billiard.

\end{document}